\documentclass[a4paper,fleqn,usenatbib]{mnras}
\usepackage{mathptmx}
\newcommand{\revise}[1]{#1}

\newcommand{\GC}{GC}%globular cluster}
\newcommand{\program}[1]{{\scriptsize {\MakeUppercase{#1}}}}% write computer programs in small capital letters

\usepackage[T1]{fontenc}
\usepackage{ae,aecompl}

\usepackage{graphicx}	% Including figure files
\usepackage{amsmath}	% Advanced maths commands
\usepackage{amssymb}	% Extra maths symbols

\title[Orbit and physical property of the NWS progenitor]{The nature of the progenitor of the M31 North-western stream: globular clusters as milestones of its orbit}
\author[T. Kirihara et al.]{
T. Kirihara,$^{1}$\thanks{E-mail: kirihara@ccs.tsukuba.ac.jp}
Y. Miki,$^{1,2}$
M. Mori,$^{1,3}$\\
$^{1}$Center for Computational Sciences, University of Tsukuba, Tennodai 1-1-1, Tsukuba, Ibaraki, 305-8577, Japan\\
$^{2}$CREST, JST, Tennodai 1-1-1, Tsukuba, Ibaraki, 305-8577, Japan\\
$^{3}$Faculty of Pure and Applied Physics, University of Tsukuba, Tennodai 1-1-1, Tsukuba, Ibaraki, 305-8577, Japan\\
}

\date{Accepted XXX. Received YYY; in original form ZZZ}

\pubyear{2016}

\begin{document}
\label{firstpage}
\pagerange{\pageref{firstpage}--\pageref{lastpage}}
\maketitle

\begin{abstract}
%  last updated on 2017/04/11(²ÐÍËÆü) 11:16:25

We examine the nature, possible orbits and physical properties of the progenitor of the North-western stellar stream (NWS) in the halo of the Andromeda galaxy (M31). 
The progenitor is assumed to be an accreting dwarf galaxy with globular clusters (GCs). 
It is, in general, difficult to determine the progenitor's orbit precisely because of many necessary parameters. 
Recently, \citet{Veljanoski2014} reported five GCs whose positions and radial velocities suggest an association with the stream.
We use this data to constrain the orbital motions of the progenitor using test-particle simulations. 
Our simulations split the orbit solutions into two branches according to whether the stream ends up in the foreground or in the background of M31. 
Upcoming observations that will determine the distance to the NWS will be able to reject one of the two branches. 
In either case, the solutions require that the pericentric radius of any possible orbit be over 2~kpc. 
We estimate the efficiency of the tidal disruption and confirm the consistency with the assumption for the progenitor being a dwarf galaxy. 
The progenitor requires the mass $\ga 2\times10^6 M_{\sun}$ and half-light radius $\ga 30$~pc. 
In addition, $N$-body simulations successfully reproduce the basic observed features of the NWS and the GCs' line-of-sight velocities. 

\end{abstract}

\begin{keywords}
galaxies: individual(M31)-galaxies: interactions%-galaxies : kinematics and dynamics.
%keyword1 -- keyword2 -- keyword3
\end{keywords}

%%%%%%%%%%%%%%%%%%%%%%%%%%%%%%%%%%%%%%%%%%%%%%%%%%

\section{Introduction}

The lambda cold dark matter ($\Lambda$CDM) theory has been successful in explaining the large-scale structure in the universe, but several discrepancies with observations have arisen below the galaxy scale. 
For example, the observed density profile of the dark matter halo (DMH) in nearby dwarf galaxies is almost flat at the centre of these galaxies, 
whereas cosmological simulations based on the $\Lambda$CDM model have generally predicted a steep power-law density distribution \citep[the ``cusp-core problem'', e.g.,][]{Moore1994,Burkert1995,Navarro1996,Ogiya2014}. 
Recently, modeling the giant southern stream in the halo of the Andromeda galaxy (M31), \citet{Kirihara2014} showed that the outer density profile of the DMH is steeper than that of the CDM prediction.

Furthermore, cosmological $N$-body simulations have predicted that CDM halos of the Milky Way (MW) and M31 size galaxies host 100--1000 sub halos but only 2--3 dozens satellite galaxies have been discovered in these galaxies \citep[the ``missing satellite (sub-halo) problem'', e.g.,][]{Moore1999,McConnachie2012}. 
The stellar debris of galaxy mergers, especially the stellar streams of satellite dwarfs, can be used to study this problem. 
If a sub-halo has little gas and/or few stars, the halo would not be detected by current telescopes. 
However, one powerful approach to confirm the existence of these invisible sub-halos is to look for gaps in stellar streams perturbed by sub-halos \citep{Carlberg2012b}. 
This technique has been employed to estimate the number of sub-halos in the MW halo \citep{Carlberg2012,Ibata2016}. 

In this work, we direct our attention to the North-western stream in M31 (hereafter NWS). This fairly long stellar stream extending over 100~kpc in projected distance in M31's halo was reported by the Pan-Andromeda Archaeological Survey (PAndAS) in \cite{McConnachie2009} and \cite{Richardson2011}. 
Despite the large scale, physical properties such as the distance and line-of-sight (LOS) velocity of the NWS have not been reported due to its faintness. 
As a result, little is known about the formation history of the NWS. 
Its LOS location relative to M31 is not known, and its total extent and mass are not well established. 
Because of the faintness, \citet{Carlberg2011} modeled a globular cluster (GC) being the progenitor. 
However, a satellite dwarf cannot be excluded. 
Only recently, \citet{Veljanoski2013} reported that a number of {\GC}s appear to spatially coincide with the stellar debris and streams in M31's halo. 
\citet{Veljanoski2014} measured the LOS velocities of these {\GC}s and found that some {\GC}s aligned with the streams exhibited a clear correlation between their positions and LOS velocities. This included five GCs aligned with the NWS. 
This result leads us to consider the possibility that the progenitor dwarf galaxies of large stellar streams carry GCs with them. 
Here, we explore progenitor orbits for the NWS, constrained by the LOS velocities of the five {\GC}s that show kinematic evidence for being associated with the stream. 

The possible orbits of the NWS progenitor are explored by a systematic parameter survey using test-particle simulations. 
In addition to the observed LOS velocities of the {\GC}s located along the NWS, we use the plane of sky position of the NWS as criteria for the likelihood of an orbit. 
In \S 2, we describe the observational data for the NWS. 
In \S 3, we outline our modelling and show the results of our test-particle simulations. 
Demonstrations with $N$--body progenitors are also covered in \S 3. 
We conclude with a brief discussion and summary in \S 4.

\section{Model and assumption}\label{sec:model}

The NWS is observed at distances of $30~{\rm kpc}<R_{\rm proj}<150~{\rm kpc}$ from the centre of M31 \citep{McConnachie2009,Lewis2013}. 
We estimated by eye the width of the stream and the location of the centre of the stream in eight regions along the length of the stream. The eight regions lie within $-5\degr.50<\xi<-2\degr.53$ and $-0\degr.50<\eta<4\degr.71$  (see \autoref{tab:table1}), where $\xi$ and $\eta$ are the sky coordinates centred on the M31 \citep{Ferguson2002}. 

\begin{table}
 \begin{minipage}{84mm}
\centering
  \caption{Position of the sampled points on the NWS. 
\label{tab:table1}}
  \begin{tabular}{cccc}
  %% \begin{tabular}{@{}cccc@{}}
  \hline
name & $\xi$ & $\eta$ & $W_{\mathrm h}$\\
 \hline
NWS-f1 & $-$2\degr.53 & $-$0\degr.50 & 0\degr.18 \\
NWS-f2 & $-$2\degr.93 & $-$0\degr.09 & 0\degr.19 \\
NWS-f3 & $-$3\degr.59 & 0\degr.58 & 0\degr.14 \\
NWS-f4 & $-$4\degr.17 & 1\degr.45 & 0\degr.16 \\
NWS-f5 & $-$4\degr.65 & 2\degr.33 & 0\degr.14\\
NWS-f6 & $-$4\degr.96 & 3\degr.17 & 0\degr.18\\
NWS-f7 & $-$5\degr.23 & 4\degr.06 & 0\degr.20\\
NWS-f8 & $-$5\degr.50 & 4\degr.71 & 0\degr.25\\\hline
\end{tabular}
\end{minipage}
\end{table}

So far, seven GCs are discovered along the NWS, and five (PAndAS-04, 09, 10, 11 and 12) of them show a clear correlation of their galactocentric radial velocities with projected radii from M31's centre \citep[see table 4 and figure 11 of][]{Veljanoski2014}. 
In this paper, we assume that these five {\GC}s initially inhabited the progenitor of the NWS and still have the same orbital motion as the NWS. 

In our simulations, we adopt an M31 model for which the gravitational potential is fixed, which reproduces well the observed surface brightness of the bulge and disc components, the velocity dispersion of the bulge, and the rotation curve of the disc \citep{Fardal2007}. 
The potential consists of a Hernquist bulge \citep{Hernquist1990}, an exponential disc and an NFW DMH \citep{Navarro1996}. 
The scale radius and total mass of the bulge are $0.61$~kpc and $3.24 \times 10^{10}M_{\sun}$, respectively. 
The M31 disc is set to have a scale height of $0.6$~kpc, a radial scale length of $5.4$~kpc, a total mass of $3.66 \times 10^{10}M_{\sun}$ and a central surface density of $2.0\times 10^8 M_{\sun}\;{\rm kpc}^{-2}$. 
The inclination and position angle of M31's disc are $77\degr$ and $37\degr$, respectively \citep{Geehan2006}. 
The scale radius and scale density of the NFW halo are $7.63$~kpc and $6.17 \times 10^7 M_{\sun} \rm{kpc}^{-3}$, respectively. 
We adopt a distance and LOS velocity to M31 of 780~kpc and $-300$ km~s$^{-1}$, respectively \citep{Font2006}. 

\section{Simulations}
\subsection{Parameter study of test-particle simulation}
\label{sec:test_particle}

To investigate possible orbits of the NWS progenitor, we perform a parameter study using test-particle simulations. The test particle represents the progenitor. 
The initial position of the test-particle is $(\xi$, $\eta)=(-4\degr.57, 2\degr.22)$ and its LOS velocity $V_{\rm los}=-472$ km~s$^{-1}$. 
These values correspond to the observed values of GC PAndAS-12, which has the smallest uncertainty in LOS velocity. 
Our results are barely affected by the fixed condition for the GC PAndAS-12 ($\xi,\eta,V_{\rm LOS}$). 
Six-dimensional phase-space information is required to specify an orbit. 
The remaining three free parameters are the distance $d$ and the proper motion ($V_{\xi}$ and $V_{\eta}$) of the NWS at the location of the initial conditions. 
We examine 5 068 617 orbit models: the parameter ranges are $-250 \;{\rm km\;s^{-1}}\leqq V_{\xi}<250\;{\rm km\;s^{-1}}$ and $-200\;{\rm km\;s^{-1}}\leqq V_{\eta} \leqq 250\;{\rm km\;s^{-1}}$ in intervals of $3$~${\rm km\;s^{-1}}$ and $500\;{\rm kpc}\leqq d \leqq 1100\;{\rm kpc}$ in intervals of $3$~kpc. 

We use the 2nd-order leap-frog method, and the time step of the integration both forwards and backwards in time is 1~kyr. 
An orbital calculation stops at 5~Gyr in each direction of integration or at the turnaround point in the sky. 
To evaluate how well the orbit reproduces the observed NWS's locus, we conduct a $\chi^2_{\nu}$ analysis for the position and LOS velocity among the simulated orbit, the observed position of the NWS and the observed LOS velocity of the five GCs. 
With respect to the loci of the NWS, we use the eight sampled regions on the NWS mentioned in \S~\ref{sec:model} (\autoref{tab:table1}). 
Observed positional uncertainties are assumed to be $0\degr.3$, which is a conservative bound based on the largest value of the estimated stream half-width $W_{\mathrm h}$ (see Table \autoref{tab:table1}). 
As for the LOS velocity, we perform the comparison with the data of the five {\GC}s at the points of closest approach in the orbit. 

\begin{figure}
  \includegraphics[width=\linewidth]{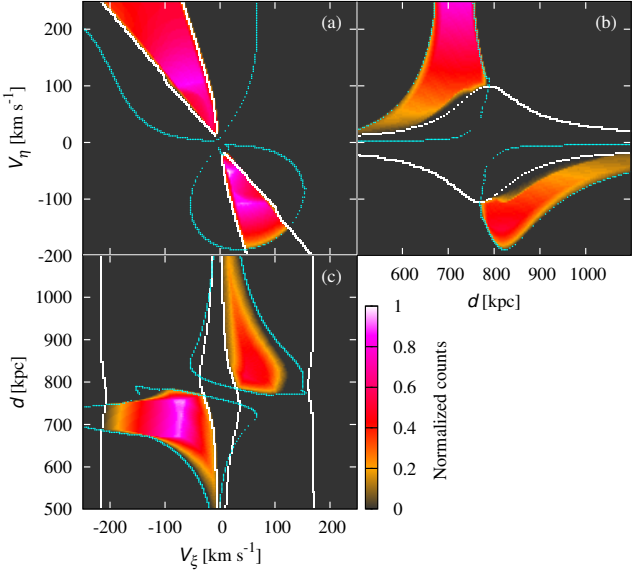}
\caption{Projected view of the possible parameter space of the NWS progenitor's orbit. 
The coloured region satisfies the 1$\sigma$ CL for both the position and the LOS velocity criteria. 
Regions excluded by each criterion are denoted by white (NWS positions) and cyan ({\GC}s' LOS velocities) dotted lines. 
A 3D plot is provided in the online-only material. 
}
\label{fig:s2plot}
\end{figure}

Fig.~\ref{fig:s2plot} illustrates the regions in parameter space yielding positions and the LOS velocities that lie within the $1\sigma$ confidence limit (CL) for both criteria. 
The $1\sigma$ CLs for the positions ($\nu=7$) and the LOS velocities ($\nu=4$) are 1.36 and 1.57, respectively and are shown in white and cyan curves, respectively. 
This figure clearly shows that the solutions that produce an NWS-like orbit consist of two main branches ($V_{\eta}>0$ and $V_{\eta}<0$) . 
The criteria for the NWS positions strongly limit the possible parameter space in the $V_{\xi}-V_{\eta}$ plane. 
The criteria for the {\GC}s' LOS velocities primarily constrain the $d-V_{\eta}$ and $V_{\xi}-d$ planes. 
A total of 102 073 orbit models satisfy both criteria, and we shall refer to this set as the set of successful orbits. A total of 615 739 and 306 968 models satisfy only the position criterion and only the LOS velocity criterion, respectively. The minimum $\chi^2_{\nu}$ values for the position and LOS velocity are 0.082 and 0.113, respectively. 

\begin{figure}
  \includegraphics[width=0.97\linewidth]{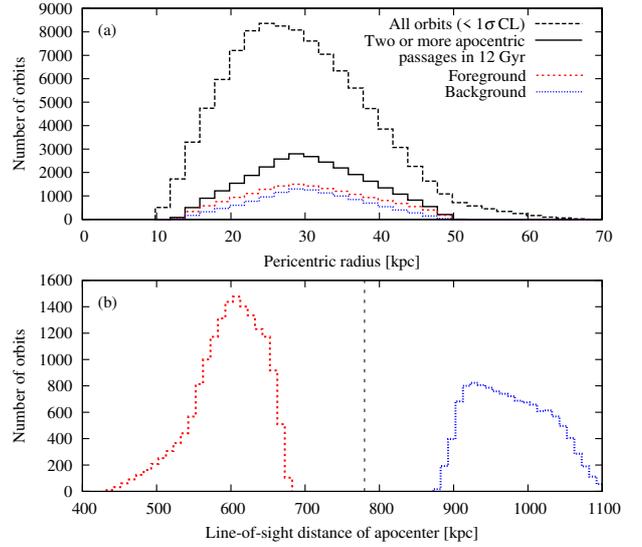}
  \caption{(a) Histogram of the pericentric radii of the successful orbits. 
The black long-dashed line shows the result for all successful orbits, which lie within the $1\sigma$ CL. 
Orbits that also have experienced two or more apocentric passages in 12~Gyr are shown with the solid black line. 
The red dashed line and the blue dotted line are the distributions of orbits whose northern apocentres are located in the foreground and background of M31, respectively. 
(b) Distributions of the apocentric distances of the orbits, which have experienced two or more apocentric passages in 12~Gyr. 
Colours correspond to the distributions in panel (a). 
The vertical dashed line marks the distance to M31 from earth. 
\label{figure1}
  }
\end{figure}

Fig.~\ref{figure1}(a) shows the distribution of the pericentric radii for all successful orbits. 
The black long-dashed line traces the distribution of such orbits. 
The orbits that have experienced two or more apocentric passages within 12~Gyr are shown with the solid black line and are divided into the two characteristic main branches, depending on whether their northern apocentres are located in the foreground or the background with respect to M31. 
The peaks of each line lie between a pericentric radius of 20~kpc and 30~kpc. 
We note that no orbit can approach M31's centre within 10~kpc. 
Even if we adopt the constraint for 3$\sigma$ CL, any orbit cannot reach 2~kpc from the M31's centre. 
Fig.~\ref{figure1}(b) describes the distribution of the apocentric distance at the northern side using the successful orbits that have experienced two or more apocentric passages in 12~Gyr. 
There clearly exist two branches: one for which the NWS extends into the foreground, and one for which it extends to the background of the M31 halo on the northern side. 
The constraints derived from the LOS velocity of the {\GC}s do not allow apocentres in the vicinity of the distance of M31's centre. 

\subsection{Mass of NWS progenitor}

The minimum pericentric radius derived by our test-particle simulations allows us to estimate the total mass of the progenitor. 
We first estimate the Hill radius that defines a tidal radius of a satellite in a gravitational potential of the host system. 
We expect that the satellite is mostly strongly tidally distorted at the pericentre in the orbital motions. 
In this case, the Hill radius will simply be determined by the mass of the satellite within the Hill radius $r_{\rm Hill}$ and the mass of M31 $M_{<{\rm peri}}$ within the pericentric radius $r_{\rm peri}$. 
Assuming a Plummer sphere for the progenitor, we obtain,
\begin{equation}
r^2_{\rm Hill}  =  \left(\frac{m_{\rm sat}}{3 M_{<{\rm peri}}}\right)^{2/3} r^2_{\rm peri} - r^2_{\rm s}, \label{eq:hill3}
\end{equation}
where $m_{\rm sat}$ and $r_{\rm s}$ are the total mass and scale radius of the Plummer sphere. 
In the case of a Plummer sphere, the half-light radius $r_{\rm h}$ is given exactly by $r_{\rm s} = (2^{2/3}-1)^{1/2} r_{\rm h} \simeq 0.77 r_{\rm h}$. 
Equation (\ref{eq:hill3}) enables us to estimate the mass of the satellite tidally disrupted by the gravitational potential of M31. 
The stripped mass $m_{\rm strip}$ is given as $m_{\rm sat} - 3 M_{<{\rm peri}} (r_{\rm Hill}/r_{\rm peri})^3$. 

\begin{figure*}
  \includegraphics[width=\linewidth]{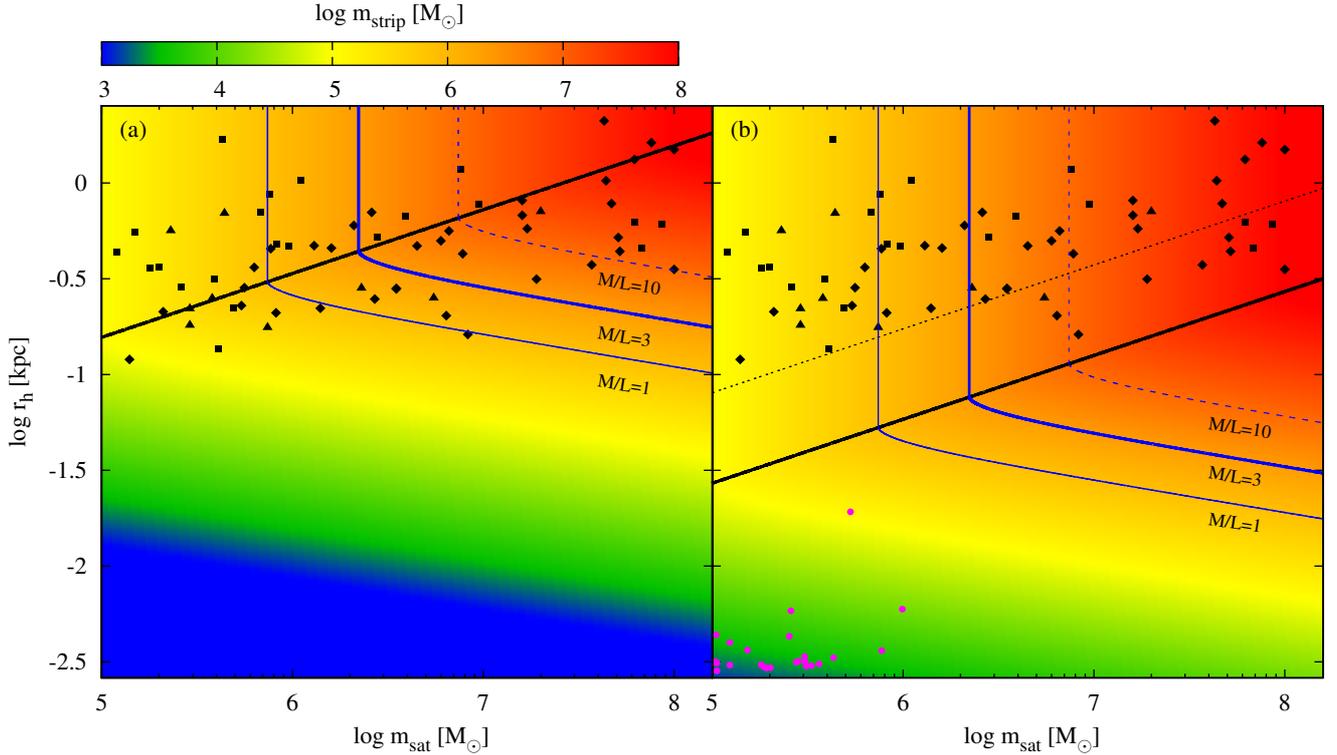}
  \caption{Stripped mass from the progenitor assuming (a) $r_{\rm peri}=25$~kpc and (b) $r_{\rm peri}=2$~kpc. 
Black triangles, squares and diamonds are dwarf galaxies in the MW, M31 and the Local Group including nearby ones, respectively \citep{McConnachie2012}. 
Magenta circles indicate {\GC}s in MW \citep{Harris1996}. 
The black lines show the critical condition for the progenitor to be completely disrupted. 
In panel b, the dotted and solid black lines show the cases of $r_{\rm peri}=10$~kpc ($1\sigma$ CL) and $r_{\rm peri}=2$~kpc ($3\sigma$ CL), respectively. 
The blue curves are minimum stripped mass from the progenitor to reproduce the estimated luminosity. 
\label{figure2}
  }
\end{figure*}

Fig.~\ref{figure2} shows the stripped mass from the progenitor as a function of the total mass and scale radius of the satellite. 
Figs.~\ref{figure2}a and \ref{figure2}b plot the results for the most probable orbit ($r_{\rm peri} = 25$~kpc) and the orbit with the smallest perigalacticon ($r_{\rm peri} = 2$~kpc), respectively. 
The black line in Fig.~\ref{figure2}a shows the critical conditions for $r_{\rm peri}= 25$~kpc, which is derived by setting $r_{\rm Hill}=0$ in equation (\ref{eq:hill3}). 
Above this line, the satellite is completely disrupted by the tidal force of the M31 gravity. 
In Fig.~\ref{figure2}, we overplot physical properties of the observed {\GC}s (magenta circles in panel b) in MW \citep{Harris1996} and dwarf galaxies (black symbols) in the Local Group including some nearby dwarf galaxies \citep{McConnachie2012}. 
To convert $M_{\rm V}$ of the {\GC}s into stellar mass, we assume a V-band mass-to-light ratio of two \citep{Pryor1993}. 
As for dwarf galaxies, the stellar masses are estimated assuming a stellar mass-to-light ratio of one \citep{McConnachie2012}. 

One might construe from the faintness of the NWS that the progenitor was a {\GC}. 
The estimation of the tidally stripped mass suggests the progenitor be a dwarf galaxy instead of a GC. 
\revise{The luminosity of the tidal stream was estimated by \citet{Carlberg2011} as $7.4\times 10^5 L_{\odot}$} using red giant branch stars obtained by the PAndAS in the metallicity range of $-2.4< {\rm [Fe/H]}<-0.6$. 
They summed over the density of the stars and corrected by the luminosity function of the GC M12, assuming their distance is equal to that of the centre of M31. 
\revise{One has to assume a $M/L$ ratio for this debris to determine the minimum mass of the progenitor. 
With a $M/L$ ratio of three, the minimum progenitor mass is $2.2\times10^6M_{\sun}$}, which is described with the blue thick solid curve in Fig.~\ref{figure2}. 
We should note that even in the case of $3\sigma$ CL, the pericentric distance must be greater than $2$~kpc. 
\revise{If the progenitor has a larger Hill radius compared to the radius of the progenitor, then it will not be disrupted and will not generate the required minimum amount of debris. 
To generate the amount of the observed debris, the progenitor must be in the upper right corner of Fig.~\ref{figure2}b. 
The limitation of the progenitor mass and half-light radius are $m_{\rm sat} \ga 2\times10^6 M_{\sun}$ and $r_{\rm h}\ga 30$~pc, respectively. 
It is not required that $r_{\rm h}$ be larger than 30~pc if the mass of the progenitor is more massive than $10^8 M_{\sun}$. 
In the mass range of GCs, the required half-light radius to create the tidal debris is too large for a GC. 
If a more likely perigalacticon of $25$~kpc is used, then the satellite mass and half-light radius must lie in the upper right corner of Fig.~\ref{figure2}a, above the $M/L=3$ curve. }

\revise{The upper right hand corners of the panels in Fig.~\ref{figure2} are inhabited by many known dwarf galaxies. 
The blue curves show, for a particular $M/L$ ratio, the mass and radius combinations that will produce the observed luminosity of the NWS, assuming that mass follows light. 
Any satellite with properties above the solid black line is expected to be completely disrupted on an orbit with perigalacticon of 25~kpc (panel a) or 2~kpc (panel b). 
For the most likely perigalacticon, 25~kpc, most of the observed satellite galaxies are near the black line, supporting the idea that the dwarf galaxy might be mostly or completely disrupted. }

The progenitor's mass can also be estimated through the metallicity of stars. 
Recently, \citet{Ibata2014b} showed that the bulk of the NWS stars have a metallicity of $-1.7< {\rm [Fe/H]} <-1.1$, which coincides with the metallicities of intermediate mass to massive ($-15 < M_{\rm V} < -10$) satellite dwarf galaxies in the Local Group \citep{McConnachie2012}. 
Using this mass-metallicity relation, the total stellar mass of the NWS progenitor is estimated to be $10^{6-8} {M_{\sun}}$. 
\revise{Therefore this result, along with the apparent coincidence of five GCs and a metallicity similar to local group dwarf galaxies, supports a consistent picture that the progenitor of the NWS is a dwarf galaxy and not a GC. }

\subsection{$N$--body simulation}

To demonstrate the tidal disruption of the NWS progenitor, we perform two $N$--body simulations representative of two branches discovered by the test-particle simulations. 
We adopt a Plummer distribution, which is constructed by the \program{MAGI} code (\citeauthor{Miki2017} in prep.) and use the gravitational octree code \program{GOTHIC} developed by \citet{Miki2016b} to run the simulations. 
The total number of particles is 65 536, the Plummer softening parameter is 16~pc and the accuracy control parameter is $2^{-7}$. 

Here, we describe the two representative models, which we label Case A and Case B. 
The physical properties and initial phase-space coordinates of the progenitor for Case A are $m_{\rm sat}=5\times10^7M_{\sun}$, $r_{\rm s}=1$~kpc and $(\xi,\eta,d,V_{\xi},V_{\eta},V_{\rm los})=(-19\degr.97,-1\degr.46, 842.38 \;{\rm kpc}, -8.53, -24.06 , -308.50)$. 
Those for Case B are $m_{\rm sat}=10^7M_{\sun}$, $r_{\rm s}=1.5$~kpc and $(\xi,\eta,d,V_{\xi},V_{\eta},V_{\rm los})=(4\degr.34, -7\degr.90,1062.49 \;{\rm kpc},-25.22, -13.42, -296.15)$. 
The units of velocity are km~s$^{-1}$. 
The pericentric distances are $r_{\rm peri}=25.54$~kpc for Case A and $r_{\rm peri}=22.11$~kpc for Case B. 

\begin{figure*}
\begin{center}
  \includegraphics[width=0.82\linewidth]{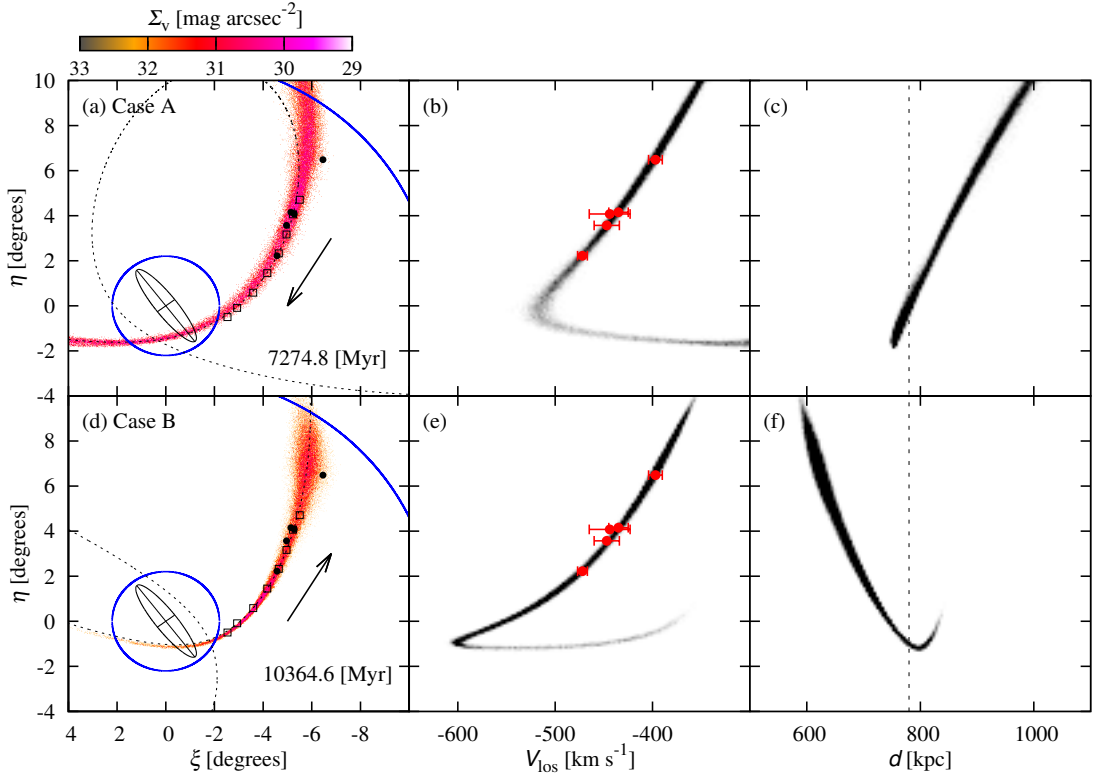}
  \caption{Spatial distribution of the simulated NWS in sky coordinates. 
Panels (a) and (d) display snapshots of the surface density for Case A and Case B, respectively. 
The streams move in the directions of the arrows. 
Test particle orbits are shown with dashed lines. 
The elapsed time of the simulation run is shown at the lower right of each panel. 
The ellipse traces M31's disc, and blue circles trace a radius of 30~kpc and 150~kpc from M31's centre. 
Filled black circles mark the positions of the observed {\GC}s aligned with the NWS. 
The open square symbols are the eight positions of the NWS used for the comparison (see \autoref{tab:table1}). 
Panels (b) and (e) show the simulated LOS velocities. 
Red circles are the observed velocities of the GCs \citet{Veljanoski2014}. 
The spatial distribution of the particles is shown in panels (c) and (f). 
The vertical dashed line corresponds to the distance of M31 from earth. 
\label{figure3}
  }
\end{center}
\end{figure*}

Fig.~\ref{figure3} displays snapshots of the results of the $N$--body simulations. 
In Case A, the progenitor moves from north to south, and the tidal debris forms a slender arch-shaped stream (see Fig.~\ref{figure3}a). 
In contrast, in Case B, the progenitor moves from south to north (Fig.~\ref{figure3}d). 
In both cases, the stream thickness becomes gradually thinner in the southern part due to M31's gravitational field. 
This change of the NWS thickness in the northern side of the M31 halo could be detected by the future observations if the contamination of the foreground dwarf stars in the MW is properly removed. 
Fig.~\ref{figure3}(b) and Fig.~\ref{figure3}(e) show that both cases match the observed LOS velocities of the GCs. 
The velocity profiles of the NWS are almost identical between two models, and the difference is only seen around the neighbourhood of the M31 disc. 
On the other hand, the stream distances are completely different in the two models. 
It is evident that the stream in Case A (Case B) lies behind (in front of) the centre of M31. 
In the future, deep photometric or spectroscopic observations will be able to distinguish between the two cases. 

%% \end{figure}

\section{Summary and Discussion}

\revise{We provide} an estimate of the total mass and phase-space information of the NWS progenitor using test-particle simulations and $N$--body simulations \revise{in a fixed M31 potential field}. 
To find the possible orbits, we use the position of the NWS and the LOS velocities of five {\GC}s thought to be associated with the NWS. 
Our test-particle simulations show that the NWS progenitor could not have passed within 2~kpc from M31's centre. 
\revise{Evaluation of the tidal force of M31 at the minimum perigalacticon gives the physical properties of the progenitor. 
Assuming a Plummer sphere as the progenitor, the minimum half-light radius is $r_{\rm h}\ga 30$~pc in the mass range of GCs, and it is too large for a GC. 
The minimum mass is set by the calculated mass of the luminous debris assuming $M/L=3$, while the minimum $r_{\rm h}$ is set by the requirement that the progenitor has enough stripped mass to account for the luminous debris. 
For the most likely perigalacticon, 25~kpc, most of the observed local group dwarf galaxies sit near the critical condition that the dwarf galaxy will be mostly or completely disrupted. 
There remains a possibility that the progenitor's central core survives. 
Since this analysis does not consider the evolution of the tidal debris, a comprehensive survey by $N$--body simulations including the effects of a progenitor's morphology, a live disc and the density profile of the DMH is necessary for further analysis \citep[cf.][]{Mori2008,Kirihara2014,Miki2016,Kirihara2017}. }

We performed $N$--body simulations of representative progenitor models, which adopt a Plummer sphere with different masses and scale lengths. 
The resulting stellar streams reproduce the position and LOS velocity of the observed {\GC}s and the shape of the observed NWS. 
Many faint stellar structures have been discovered in the M31 halo, and they often contain GCs \citep{Veljanoski2013}. 
The approach described in this paper using the LOS velocity of the GCs will be a powerful tool to explore their formation histories. 

Depending on the orbit of the progenitor, the stream at the present-day epoch lies either in front of or behind the centre of M31.
Forthcoming observational estimates of the distance of the NWS will allow us to differentiate between the two cases. 
In addition to this, a large-scale parameter study using $N$-body simulations is needed to determine the precise orbital parameters, physical properties of the progenitor and the current position of its core. 
The NWS will then become a new laboratory to examine the shape and the outer density profile of the DMH in M31 \citep{Hayashi2014,Kirihara2014}, and to search for sub-DMHs within \citep{Carlberg2012b}. 

\section*{Acknowledgements}

We thank Y. Komiyama, M. Chiba, R. Takahashi and M. Abe for fruitful discussions and appreciate A.Y. Wagner for a careful reading and comments. 
We are also grateful for the anonymous referee for fruitful and helpful suggestions. 
This work was supported by Grant--in--Aid for JSPS Fellows (T.K. 26.348) and for Scientific Research (25400222) and by the Japan Science and Technology Agency's (JST) CREST program entitled ``Research and Development of Unified Environment on Accelerated Computing and Interconnection for Post-Petascale Era''. 
Numerical computations are carried out on the HA-PACS System in the Center for Computational Sciences, University of Tsukuba, Japan. 

%%%%%%%%%%%%%%%%%%%%%%%%%%%%%%%%%%%%%%%%%%%%%%%%%%



\bibliographystyle{mn2e}%mnras}
\bibliography{kirihara}


%%%%%%%%%%%%%%%%%%%%%%%%%%%%%%%%%%%%%%%%%%%%%%%%%%

%%%%%%%%%%%%%%%%% APPENDICES %%%%%%%%%%%%%%%%%%%%%
%%\clearpage

%% %% \section{Some extra material}

%% %% If you want to present additional material which would interrupt the flow of the main paper,
%% %% it can be placed in an Appendix which appears after the list of references.

% Don't change these lines
\bsp	% typesetting comment
\label{lastpage}
\end{document}